\documentclass[tradiabstract]{aa}

\newcommand{\bea}{\begin{eqnarray}}
\newcommand{\eea}{\end{eqnarray}}
\newcommand{\be}{\begin{equation}}
\newcommand{\ee}{\end{equation}}
\newcommand{\bc}{\begin{center}}
\newcommand{\ec}{\end{center}}
\newcommand{\ben}{\begin{enumerate}}
\newcommand{\een}{\end{enumerate}}
\newcommand{\bd}{\begin{description}}
\newcommand{\ed}{\end{description}}
\newcommand{\bmi}[1]{\begin{minipage}{#1 cm}}
\newcommand{\emi}{\end{minipage}}
\newcommand{\bmif}[1]{\begin{minipage}{#1\textwidth}}

\def\llabel#1{\label{sc:#1}}
\def\elabel#1{\label{eq:#1}}

\def\eck#1{\left\lbrack #1 \right\rbrack}

\def\rund#1{\left( #1 \right)}

\def\wave#1{\left\lbrace #1 \right\rbrace}

\def\A{{\cal A}}
\def\B{{\cal B}}

\def\d{{\rm d}}

\def\arcsecf {\hbox{$.\!\!^{\prime\prime}$}}

\def\vp{\varphi}
\def\vt{{\vartheta}}

\def\Real{{\rm I\mathchoice{\kern-0.70mm}{\kern-0.70mm}{\kern-0.65mm}%
  {\kern-0.50mm}R}}
\def\C{\rm C\kern-.42em\vrule width.03em height.58em depth-.02em
       \kern.4em}

\def\bx#1{\leavevmode\thinspace\hbox{\vrule\vtop{\vbox{\hrule\kern1pt
        \hbox{\vphantom{\tt/}\thinspace{\bf#1}\thinspace}}
      \kern1pt\hrule}\vrule}\thinspace}

\def\vc#1{{\mbox{\boldmath$#1$\unboldmath}}}
{\catcode`\@=11
\gdef\SchlangeUnter#1#2{\lower2pt\vbox{\baselineskip 0pt \lineskip0pt
  \ialign{$\m@th#1\hfil##\hfil$\crcr#2\crcr\sim\crcr}}}
}

\def\ueber#1#2{{\setbox0=\hbox{$#1$}%
  \setbox1=\hbox to\wd0{\hss$\scriptscriptstyle #2$\hss}%
  \offinterlineskip
  \vbox{\box1\kern0.4mm\box0}}{}}

\def\bx#1{\leavevmode\thinspace\hbox{\vrule\vtop{\vbox{\hrule\kern1pt
        \hbox{\vphantom{\tt/}\thinspace{\bf#1}\thinspace}}
      \kern1pt\hrule}\vrule}\thinspace}

\def\arcsecf {\hbox{$.\!\!^{\prime\prime}$}}

{\catcode`\@=11
\gdef\SchlangeUnter#1#2{\lower2pt\vbox{\baselineskip 0pt \lineskip0pt
  \ialign{$\m@th#1\hfil##\hfil$\crcr#2\crcr\sim\crcr}}}
}

\def\ts{\thinspace}

\usepackage{graphicx}
\usepackage{txfonts}
\usepackage{natbib}
\usepackage{subfigure}
\usepackage{enumerate}
\usepackage{slashbox}
\begin{document}

   \title{Source-position transformation -- an approximate invariance
     in strong gravitational lensing}

   \author{Peter Schneider \inst{1} \& Dominique Sluse
          \inst{1}
          }

   \institute{Argelander-Institut f\"ur Astronomie, Universit\"at
     Bonn, Auf dem H\"ugel 71, D-53121 Bonn, Germany\\
    peter, dsluse@astro.uni-bonn.de}

%   \date{Received September 15, 1996; accepted March 16, 1997}

% \abstract{}{}{}{}{} 
% 5 {} token are mandatory
 
  \abstract
  {The main obstacle for gravitational lensing to determine accurate
    masses of deflectors, or to determine precise estimates for the
    Hubble constant, is the degeneracy of lensing observables with
    respect to the mass-sheet transformation (MST). The MST is a
    global modification of the mass distribution which leaves all
    image positions, shapes and flux ratios invariant, but which
    changes the time delay. Here we show that another global
    transformation of lensing mass distributions exists which almost
    leaves image positions and flux ratios invariant, and of which the
    MST is a special case. Whereas for axi-symmetric lenses this
    source position transformation exactly reproduces all strong
    lensing observables, it does so only approximately for more
    general lens situations. We provide crude estimates for the
    accuracy with which the transformed mass distribution can
    reproduce the same image positions as the original lens model, and
    present an illustrative example of its performance. This new
    invariance transformation most likely is the reason why the same
    strong lensing information can be accounted for with rather different mass
    models.  }

  \titlerunning{Source-position transformation}

   \maketitle
%
%________________________________________________________________

\section{\llabel{Sc1}Introduction}
Multiple image systems in strong gravitational lensing systems provide
an invaluable tool for the determination of mass properties of cosmic
objects, specifically of galaxies and galaxy clusters (see, e.g.,
Kochanek \citeyear{SaasFee3};
Bartelmann \citeyear{Bart10} and 
references therein). The determination of the mass inside the Einstein
radius of a multiple image system is the most accurate mass
measurement available for galaxies and cluster cores.

Mass estimates at larger and smaller radii are, however, less
accurate. The reason for this is that a given system of multiple
images can be fitted by more than one mass model -- i.e., the mass
model is not unique. For example, a four-image system provide a total
of six positional constraints on the lensing mass distribution, and
many different density profiles can satisfy these constraints.  The
variety of mass distribution which can reproduce a set of lensed
images can be seen by adding angular structures to the lens potential
\citep{TWH00, Evans2003}, or by modelling the mass distribution
with a grid of variable pixels or a sum of basis functions
\citep[e.g.][]{Saha1997a, Diego2005a, Coe08,
  Liesenborgs2012}. Therefore, a finite set of individual lensed
compact images clearly cannot uniquely determine the lensing mass
distribution.

When extended source components are lensed, e.g., into a partial or
full Einstein ring, the constraints on the lens model become
considerably stronger. Here, a point-by-point modification of a mass
model (like in the LensPerfect code of Coe et al. \citeyear{Coe08}) 
no longer can be used to fit the observed brightness profile
with a lens model. However, as was pointed out first by \cite{FGS85},
even in this case the mass model is not unique; there exists a
transformation of the mass distribution, called mass-sheet
transformation (MST), which leaves all image positions and image flux
ratios invariant. If $\kappa(\vc\theta)$ denotes the dimensionless
surface mass density of the lens at angular position $\vc\theta$, then
the whole family of mass models
\be
\kappa_\lambda(\vc\theta)=\lambda\kappa(\vc\theta)+(1-\lambda)
\elabel{MST}
\ee
predicts the same imaging properties as the original mass profile
$\kappa(\vc\theta)$. The MST keeps the mass inside the Einstein radius
invariant, but changes the enclosed mass at all other
radii. Furthermore, the MST changes the predicted product of time
delay between images and the Hubble constant, $\tau=H_0\,\Delta t$, to
$\tau_\lambda=\lambda \tau$, for all pairs of images. As we pointed
out in Schneider \& Sluse (\citeyear{SS13}; hereafter SS13), this MST
may strongly impact on the ability to use time-delay lens systems for
accurate determinations of the Hubble constant.

In SS13, we furthermore considered an illustrative case where a
composite lens, consisting of a Hernquist profile to resemble the
distribution of stellar mass in a lens galaxy, plus a modified
Navarro, Frank and White profile for the description of the dark
matter in the inner part of the galaxy, yields almost identical
imaging properties as a power-law mass profile. The relation between
these two mass models is not described by a MST; in particular, we
found that the time delay ratios of image pairs between these two
models are not constant, as would be predicted from a MST. It thus
appeared as if there exists a more general transformation between
lensing mass models which leaves observed image positions almost
invariant. Hints of the existence of such a transformation were
pointed out earlier by several authors (Saha and Williams
~\citeyear{Saha2006c}; Read et al. \citeyear{Read2007}, their Appendix
A3; Coe et al. \citeyear{Coe08}, their Sect.\ts 3.4), but to our
knowledge it has never been identified as a transformation of the
source plane or derived explicitly.

In this paper, we will show the existence of such a transformation,
i.e., a transformation of the deflection law which leaves the strong
lensing properties invariant not only for a finite set of source
positions, but for all source positions at the same time; for reasons
that will become obvious in the following, we call it the
source-position transformation (SPT). The general concept of the SPT
is outlined in Sect.\ts\ref{sc:Sc2}. We will then show in
Sect.\ts\ref{sc:Sc3} that for axi-symmetric lenses, the SPT is indeed
an exact invariance transformation which leaves all relative image
positions and flux ratios invariant. Thus, there exists a much larger
set of mass models than described by the MST which lead to the same
strong lensing predictions as the original mass distribution. We then
turn to the more general case in Sect.\ts\ref{sc:Sc4} and show that
the lens models obtained through an SPT in general lead to different
imaging properties, but that these differences can be quite small in
realistic cases. We consider the same example as that in SS13 to show
how the SPT works in practice; a more detailed investigation of the
SPT will be deferred to a later publication. We briefly discuss our
findings and conclude in Sect.\ts\ref{sc:Sc5}; in particular we will
discuss the issue that the SPT only yields an approximate invariance
transformation, and its relevance to application in strong lensing
systems.

\section{\llabel{Sc2}The principle of the source position
  transformation} 
A given mass distribution $\kappa(\vc\theta)$ defines a mapping from
the lens plane $\vc\theta$ to the source plane,
$\vc\beta=\vc\theta-\vc\alpha(\vc\theta)$; throughout this paper, we
use standard gravitational lensing notation \citep[see,
e.g.,][]{SaasFee1}.  Provided the mass is sufficiently concentrated,
there will be regions in the source plane such that if a source is located
there, it has multiple images, that is, several points $\vc\theta_i$
correspond to the same source position. The source position
corresponding to these images is not observable; hence, the constraint
imposed on the lens from observing $n$ such multiple images is
\be
\vc\theta_i-\vc\alpha(\vc\theta_i)=\vc\theta_j-\vc\alpha(\vc\theta_j)\;,
\elabel{ta}
\ee
for all $1\le i<j\le n$.  Hence, the constraints we obtain from
observing a strong lensing system is at best a relation between points
corresponding to the same source position -- i.e., a mapping
$\vc\theta_i(\vc\theta_1)$, $i\ge 2$, for all images $i$ corresponding
to the same source position as $\vc\theta_1$. Images corresponding to
singly-imaged source location carry no strong lensing information
about the lens mapping.

We now ask whether there exists another deflection law
$\hat{\vc\alpha}(\vc\theta)$ which yields exactly the same mapping
$\vc\theta_i(\vc\theta_1)$ as the original one. If such a mass
distribution exists, then the condition
\be
\vc\theta_i-\hat{\vc\alpha}(\vc\theta_i)
=\vc\theta_1-\hat{\vc\alpha}(\vc\theta_1)
\elabel{tahat}
\ee
must be satisfied, for all images $i$ ($\ge 2$) corresponding to the
same source position $\vc\beta$ as $\vc\theta_1$.

The above consideration shows that the new deflection law $\hat{\vc
  \alpha}(\vc\theta)$ provides the same mapping
$\vc\theta_i(\vc\theta_1)$ as the original one if (1) all image pairs
$\vc\theta_1$, $\vc\theta_2$ that belong to the same source position
in the original mapping are also multiple images under the new
deflection law, and (2) any two points $\vc\theta_1$, $\vc\theta_2$
which do not correspond to the same source position in the original
mapping are also not matched by the new one. Two deflection laws which
satisfy this condition are called `equivalent'. 

If such an equivalent deflection law to $\vc\alpha$ indeed exists, then the
new deflection $\hat{\vc\alpha}$ defines a new lens mapping
\be
\hat{\vc\beta}=\vc\theta-\hat{\vc\alpha}(\vc\theta)
\elabel{newlens}
\ee
from the lens plane to the source plane. The different images
$\vc\theta_i$ corresponding to the same source position $\vc\beta$
must also have the same source position $\hat{\vc\beta}$ in the new
mapping, according to (\ref{eq:tahat}). Therefore, the new deflection
law $\hat{\vc\alpha}$ defines a mapping $\hat{\vc\beta}(\vc\beta)$,
implicitly given by
\be
\vc\theta=\hat{\vc\alpha}(\vc\theta)+\hat{\vc\beta}
=\vc\alpha(\vc\theta)+\vc\beta\;,
\elabel{N4}
\ee
where $\vc\theta$ is any of the possible multiple images corresponding
to the source position $\vc\beta$.\footnote{Note that for any source
  position $\vc\beta$, at least one image $\vc\theta$ exists,
  according to the odd-number theorem~\citep{Burke1981}.}

We can reverse the argument and consider a mapping
$\hat{\vc\beta}(\vc\beta)$ from the original source coordinates to the
new ones; such a mapping in the source plane gives rise to a modified
deflection law, as seen by (\ref{eq:N4}),
\be
\hat{\vc\alpha}(\vc\theta)=\vc\alpha(\vc\theta)
+\vc\beta-\hat{\vc\beta}
=\vc\theta-\hat{\vc\beta}(\vc\theta-\vc\alpha(\vc\theta))\;,
\elabel{hatalpha}
\ee
where in the last step we inserted the original lens equation
$\vc\beta=\vc\theta-\vc\alpha(\vc\theta)$. Therefore, any
source-position transformation (SPT) $\hat{\vc\beta}(\vc\beta)$
defines a deflection law $\hat{\vc\alpha}(\vc\theta)$ 
such that all images of the same source under the original lens
mapping are also multiple images with the new lens equation
(\ref{eq:newlens}). Thus, the
two lens mappings caused by $\vc\alpha$ and $\hat{\vc\alpha}$ predict
the same multiple images, for all source positions $\vc\beta$ (or
$\hat{\vc\beta}$).

The mass-sheet transformation (MST) is a special case of this more
general SPT, obtained by setting $\hat{\vc\beta}=\lambda
\vc\beta$, which gives rise to the transformed deflection law
\be
\hat{\vc\alpha}(\vc\theta)
=\vc\theta-\lambda[\vc\theta-\vc\alpha(\vc\theta)]
=\lambda\vc\alpha(\vc\theta)+(1-\lambda)\vc\theta\;,
\elabel{hatalphaMST}
\ee
which we recognize being the deflection of the mass-sheet transformed
deflection $\vc\alpha$, corresponding to the transformed covergence
$\kappa_\lambda$ in (\ref{eq:MST}).

The Jacobi matrix of the new lens equation (\ref{eq:newlens}) reads
\be
\hat\A(\vc\theta) ={\partial \hat{\vc\beta}\over\partial\vc\theta}
 ={\partial \hat{\vc\beta}\over\partial\vc\beta}
{\partial {\vc\beta}\over\partial\vc\theta}\equiv{\cal
  B}(\vc\beta(\vc\theta))\,\A(\vc\theta)\;,
\elabel{hatA}
\ee
where ${\cal B}$ is the Jacobi matrix of the SPT and $\A$ the Jacobi
matrix of the original lens equation. This implies that
\be
\det{\hat\A}=\det{\cal B}\, \det\A\;.
\ee
Hence, if the SPT $\hat{\vc\beta}(\vc\beta)$ is a one-to-one mapping
(with no loss of generality, we will require
$\det{\cal B}>0$ for all $\vc\beta$), the critical curves of the
modified lens mapping are exactly the same as those of the original
lens mapping.\footnote{If the mapping $\hat{\vc\beta}(\vc\beta)$ is
  not one-to-one, then there exist pairs of positions $\vc\beta^{(1)}$
  and $\vc\beta^{(2)}$ which are mapped onto the same
  $\hat{\vc\beta}$. This implies that all images $\vc\theta_i^{(1)}$
  and $\vc\theta_i^{(2)}$ which correspond to these two different source
  positions are images of the same source $\hat{\vc\beta}$ in the new
  mapping, modifying the pairing of images. Hence, we will assume
  $\det{\cal B}>0$ in the following.} 

From (\ref{eq:hatA}), we infer that the relative magnification
matrices between image pairs from the same source $\hat{\vc\beta}$
remain unchanged,
\be
\hat\A(\vc\theta_1)\hat\A^{-1}(\vc\theta_2)
=\A(\vc\theta_1)\A^{-1}(\vc\theta_2)\;,
\ee
which implies that magnification ratios of image pairs are preserved,
as well as their relative image shapes. 

To summarize this section: Any bijective SPT
$\hat{\vc\beta}(\vc\beta)$ (with $\det{\cal B}>0$) defines an
equivalent deflection law $\hat{\vc\alpha}(\vc\theta)$ given by
(\ref{eq:hatalpha}), i.e., which yields the same strong lensing
properties as the original lens mapping. However, this does in general
not imply that there is a corresponding mass distribution
$\hat\kappa(\vc\theta)$ which yields the deflection law
$\hat{\vc\alpha}$, owing to the fact that in general, the Jacobian
matrix $\hat\A$ will be non-symmetric (and thus the deflection
$\hat{\vc\alpha}$ cannot be derived as a gradient of a deflection
potential); we will discuss this issue in more detail in
Sect.\ts\ref{sc:Sc4}. However, for the special case of axi-symmetric
lenses, such modified mass distributions do exist, as discussed next.

\section{\llabel{Sc3}The axi-symmetric case}
We first consider the case of an axi-symmetric lens, for which the SPT
yields an exact invariance transformation between different mass
profiles $\kappa(\theta)$, which will be explicitly derived in
Sect.\ts\ref{sc:Sc3.1}. In Sect.\ts\ref{sc:Sc3.2}, we provide a few
examples for such modified density profiles.

\subsection{\llabel{Sc3.1}The SPT-transformed mass profile}
Let $\kappa(\theta)$ be the radial
mass profile of a lens, corresponding to the lens mapping
$\beta=\theta-\alpha(\theta)$, and let
\be
\hat\beta=\eck{1+f(\beta)}\beta
\elabel{1D-SPT}
\ee
describe the SPT. To preserve axi-symmetry, the deformation function
$f(\beta)$ must be even, $f(-\beta)=f(\beta)$. Furthermore, we require
the SPT (\ref{eq:1D-SPT}) to be one-to-one, i.e.,
$\hat\beta'=1+f+\beta f'>0$.  The modified deflection law is,
according to (\ref{eq:hatalpha}),
\bea
\hat\alpha(\theta)&=&\theta-\eck{1+f(\theta-\alpha(\theta))}
[\theta-\alpha(\theta)]\nonumber \\
&=&\alpha(\theta)-f(\theta-\alpha(\theta))[\theta-\alpha(\theta)]\;.
\eea
We will show next that this deflection can be derived from a mass
profile $\hat\kappa(\theta)$. For this, we first write the deflection
as $\hat\alpha(\theta)=\hat m(\theta)/\theta$, where 
\be
\hat m(\theta)=2\int_0^\theta \d\theta'\;\theta'\,\hat\kappa(\theta')
\ee
is the enclosed dimensionless mass within $\theta$. This yields
\be
\hat m(\theta)=\theta\,\hat\alpha(\theta)
=\theta\,\alpha(\theta)-\theta\eck{\theta-\alpha(\theta)}
f(\theta-\alpha(\theta))\;.
\ee
The mass profile $\hat\kappa$ is obtained from $\hat m$ through
$\hat\kappa(\theta)=\hat m'(\theta)/(2\theta)$; calculating the
derivative, we find
\bea
\hat m' &=& \theta\, \alpha'+\alpha - \rund{2\theta-\alpha-\theta\alpha'}f
-\theta(\theta-\alpha)  \rund{1-\alpha'} f' \nonumber \\
&=&
2\theta\, \kappa - 2\theta(1-\kappa)f-\theta^2\det\A\, f'\;,
\eea
where we now dropped the arguments of the functions, keeping in mind
that $\alpha$ depends on $\theta$ and $f$ on $\beta=\theta-\alpha$;
furthermore, we used that $\d\beta/\d\theta=1-\alpha'$, and in the
last step, we employed the relations $\alpha'+\alpha/\theta=2\kappa$
and $\det\A=(1-\alpha/\theta)(1-\alpha')$ which apply for the
axi-symmetric case.\footnote{Since $\alpha=m/\theta$, 
$\alpha'=m'/\theta-m/\theta^2=2\kappa-\alpha/\theta$ and 
$\det\A=(\beta/\theta)(\d\beta/\d\theta)$.} Hence,
\bea
\hat \kappa(\theta)={\hat m'(\theta)\over 2\theta}
&=&\kappa(\theta)-[1-\kappa(\theta)]\,f(\theta-\alpha(\theta)) \nonumber \\
&-&{\theta\over 2}
\det\A(\theta)\,f'(\theta-\alpha(\theta))\;.
\elabel{kh-axi}
\eea
This equation now yields an explicit expression for the mass profile
$\hat\kappa(\theta)$ of the transformed lens mapping, in terms of the
original mass distribution $\kappa(\theta)$ and the source-plane
deformation $f(\theta)$. The mass profiles $\kappa(\theta)$ and
$\hat\kappa(\theta)$ thus predict exactly the same lensing properties
concerning multiple images and flux ratios of compact images, as well
as the multiple images of extended source components. Whereas the
magnifications are affected, as well as the corresponding shapes of
the sources, these properties are unobservable in general in strong
lensing systems.  

\subsection{\llabel{Sc3.2}Some examples of transformed density
  profiles} 
In order to get more insight into this transformation, we first consider
some special locations in the lens plane, starting with the
tangential critical curve at $\theta=\theta_{\rm E}$, where
$\alpha(\theta_{\rm E})=\theta_{\rm E}$ and thus $\beta=0$. Writing
$\theta \det\A=(\theta-\alpha)(1-\alpha')$ and using
$\alpha'=2\kappa-\alpha/\theta$, we find from differentiating
(\ref{eq:kh-axi}) that at the Einstein radius
\bea
\hat\kappa(\theta_{\rm E})&=&\kappa(\theta_{\rm E})\eck{1+f(0)}
-f(0)\;, \nonumber\\ 
\hat\kappa'(\theta_{\rm E})&=&
\kappa'(\theta_{\rm E})\eck{1+f(0)}\;, \nonumber\\
\hat\kappa''(\theta_{\rm E})&=&
\kappa''(\theta_{\rm E})\eck{1+f(0)}
-12\eck{1-\kappa(\theta_{\rm E})}^3 f''(0)\;, \elabel{17}\\
\hat\kappa'''(\theta_{\rm E})&=&
\kappa'''(\theta_{\rm E})\eck{1+f(0)}
+12\eck{1-\kappa(\theta_{\rm E})}^2 f''(0) \nonumber \\
&&\times
\wave{6\kappa'(\theta_{\rm E})+{5\eck{1-\kappa(\theta_{\rm
        E})}\over\theta_{\rm E}}}\;,\nonumber
\eea
where we also used the fact that $f$ is an even function, i.e., its
odd derivatives vanish at the origin.
These relations show how the mass profile can be modified near
the Einstein radius. We can choose $\hat\kappa(\theta_{\rm E})$ freely with an
appropriate choice of $f(0)$, but that fixes the slope of
$\hat\kappa$ at $\theta_{\rm E}$, which is as well determined by
$f(0)$. This connection between $\hat\kappa(\theta_{\rm E})$ and
$\hat\kappa'(\theta_{\rm E})$ is the same as for the MST. The new
feature of the SPT shows up
for the curvature of the mass profile at $\theta_{\rm E}$ for which we
again can make a choice, but then the third derivative is fixed -- and
so on. The fact that $f$ is an even function implies that we can make
a choice for all even derivatives, but the odd derivatives are then
tied to the former.

Choosing the derivatives of the mass profile at the Einstein radius by
selecting $f(0)$ and its even derivatives then fixes the expansion of
the central surface mass density through
\bea
\hat\kappa(0)&=& \kappa(0) \eck{1+f(0)} -f(0) \;,\nonumber \\
\hat\kappa'(0)&=& \kappa'(0) \eck{1+f(0)}  \;,\nonumber \\
\hat\kappa''(0)&=& \kappa''(0) \eck{1+f(0)} -2\eck{1-\kappa(0)}^3 f''(0) \;, \\
\hat\kappa'''(0)&=& \kappa'''(0) \eck{1+f(0)}
+15\eck{1-\kappa(0)}^2\kappa'(0) f''(0) \;. \nonumber 
\eea
In particular, if the mass profile is smooth at the origin, so that
all odd derivatives of $\kappa$ vanish there, the same property will
be shared by the transformed mass distribution.

If there is a radial critical curve, which is the case if
$\kappa(\theta)$ is a regular function, then at $\theta_{\rm c}$,
$1-\alpha'(\theta_{\rm c})=0$. The corresponding caustic in the source
plane has radius $\beta_{\rm c}=\alpha(\theta_{\rm c})-\theta_{\rm
  c}$, with $\alpha(\theta_{\rm c})=\theta_{\rm
  c}\eck{2\kappa(\theta_{\rm c})-1}$. At this location, we then obtain
\bea
\hat\kappa(\theta_{\rm c})&=&\kappa(\theta_{\rm
  c})\eck{1+f(\beta_{\rm c})}-f(\beta_{\rm c}) \nonumber \\
\hat\kappa'(\theta_{\rm c})&=&\kappa'(\theta_{\rm
  c})\eck{1+f(\beta_{\rm c})}
+2\eck{1-\kappa(\theta_{\rm c})} f'(\beta_{\rm c}) \nonumber \\
&&\qquad \times
\eck{\kappa(\theta_{\rm c})+\theta_{\rm c}\kappa'(\theta_{\rm c})-1}
\\
\hat\kappa''(\theta_{\rm c})&=&\kappa''(\theta_{\rm
  c})\eck{1+f(\beta_{\rm c})}
+2\eck{1-\kappa(\theta_{\rm c})} f'(\beta_{\rm c})\nonumber \\
&&\qquad \times
\eck{2-2\kappa(\theta_{\rm c})+\theta_{\rm c}^2\kappa''(\theta_{\rm
    c})}/\theta_{\rm c} \;. \nonumber 
\eea

\subsection{Examples}
Our first set of examples for the action of the SPT is constructed
such that near the tangential critical curve, the transformed mass
distribution is approximately a power law, $\hat\kappa(\theta)
\approx \hat\kappa(\theta_{\rm E}) (\theta/\theta_{\rm E})^{-\nu}$,
for $\theta$ close to $\theta_{\rm E}$. For such a power-law
distribution, one finds that 
\be
\nu=-{\hat\kappa'(\theta_{\rm E})\,\theta_{\rm E}\over
  \hat\kappa(\theta_{\rm E})}\;, \quad
\theta_{\rm E}^2 \hat\kappa''(\theta_{\rm E})
=\rund{{\hat\kappa'(\theta_{\rm E})\,\theta_{\rm E}\over 
\hat\kappa(\theta_{\rm E})} -1}
\hat\kappa'(\theta_{\rm E})\,\theta_{\rm E} \;.
\elabel{PLC}
\ee
If we choose a singular isothermal sphere (SIS) as the original mass
model, with $\kappa=\theta_{\rm E}/(2\theta)$, $\alpha=\theta_{\rm
  E}$, and use the first three relations of (\ref{eq:17}), the second
equation of (\ref{eq:PLC}) then yields a condition for the second
derivative of $f$ at the origin,
\be
 \theta_{\rm E}^2 f_2=-{2 f_0 (1+f_0)\over 3 (1-f_0)}\;,
\ee
where $f_0\equiv f(0)$, $f_2\equiv f''(0)$, and the local slope 
is $\nu=(1+f_0)/(1-f_0)$. 
\begin{figure}
\includegraphics{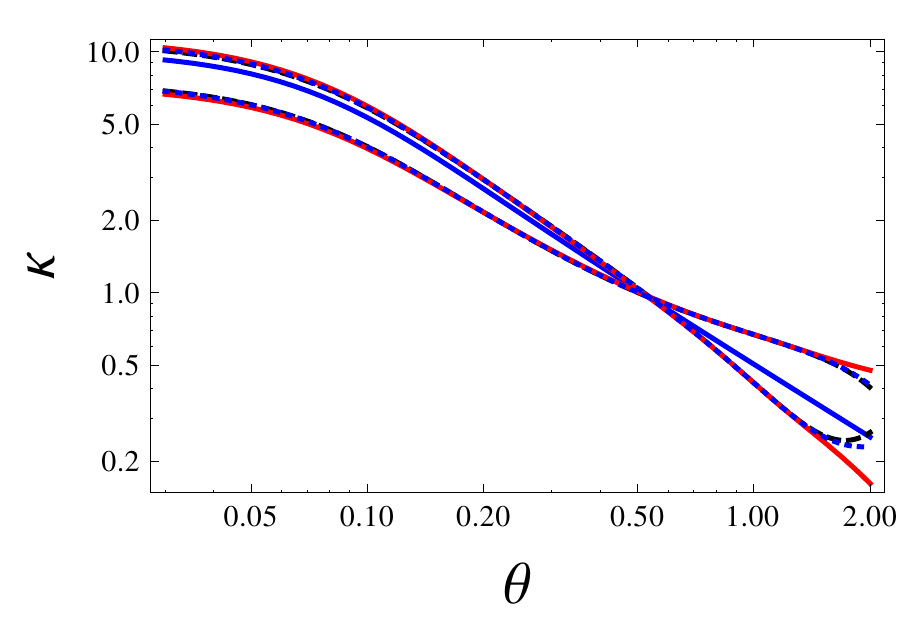}
\caption{Axi-symmetric example of an SPT between a non-singular
  isothermal sphere (blue solid curve), with core $\theta_{\rm
    c}=0.1\theta_{\rm E}$, and other mass profiles. The other curves
  are transformed mass profiles, using the SPT, with different
  deformation function $f(\beta)$, all satisfying the power-law
  condition (\ref{eq:PLC}), and with slope $\nu=0.5$ (flatter curves)
  and $\nu=1.4$ (steeper curves): The dashed black curves correspond
  to the polynomial $f(\beta)=f_0+f_2 \beta^2/2$, the red curves to
  $f(\beta)=f_0+\beta_0^2 f_2 \beta^2/[2(\beta^2+\beta_0^2)]$, with
  $\beta_0=0.8 \theta_{\rm E}$, and the blue dotted curves to
  $f(\beta)=2f_0/\cosh(\beta/\beta_0)-f_0$, with $\beta_0= \theta_{\rm
    E}\sqrt{3(1-f_0)/(1+f_0)}$}
\label{fig:Fig1}
\end{figure}
In Fig.\ts\ref{fig:Fig1} we have plotted the original mass profile,
where we have taken a non-singular isothermal sphere with core radius
$\theta_{\rm c}=0.1\theta_{\rm E}$; the introduction of a small core
does affect the foregoing relations only very little, as these are
obtained by considering $\kappa(\theta)$ at the Einstein radius.  For
two values of the slope $\nu$ near the Einstein radius, $\nu=0.5$ and
$\nu=1.4$, we have plotted three different transformed mass profiles,
where the corresponding functions $f(\beta)$ are described in the
figure caption. For all these cases, the local behavior near the
Einstein radius is indeed well approximated by a power law. Some of
the profiles become unphysical near $\theta\sim 2$, i.e., $\beta\sim
1$, which is due to the fact that the first derivative of $f$ which
enters (\ref{eq:kh-axi}) becomes too large there for the corresponding
deformation function. Nevertheless, these simple example already show
the range of freedom the SPT offers to generate axi-symmetric mass
profiles with identical strong lensing behavior. We also point out
that we plotted the mass profiles only up to $\theta=2\theta_{\rm E}$,
i.e., in the angular range where multiple images occur. For larger
$\theta$, the mass profile can be chosen arbitrarily, without
constraints.

We next consider the case of large $\beta>\beta_{\rm c}$, for
which no multiple images occur, so that the lens mapping becomes
one-to-one there. Hence, for sufficiently large $\theta$, the lens
equation defines a mapping $\theta(\beta)$. We then can rewrite
(\ref{eq:kh-axi}) in the form
\be
f'(\beta)+{2[1-\kappa(\theta)]\over \theta\,\det\A(\theta)}f(\beta)
={2\eck{\kappa(\theta)-\hat\kappa(\theta)}\over
  \theta\,\det\A(\theta)}\;,
\ee
or by using the mapping $\theta(\beta)$, 
\be
f'(\beta)+g(\beta)f(\beta)=h(\beta)\;,
\ee 
where $g=2(1-\kappa)/(\theta\,\det\A)$ and
$h=2(\kappa-\hat\kappa)/(\theta\,\det\A)$ are functions of $\beta$.
For a given $\kappa$ and a target $\hat\kappa$, the function $f$ can
be determined by solving this differential equation.

We will give a simple example for this procedure. Assume $\kappa$ to
describe an SIS. Furthermore, let our target density profile be
$\hat\kappa(\theta)= (\theta_1 /\theta)^2$. This yields 
\be
g={2(1-\theta_{\rm E}/2 \theta)\over(\theta-\theta_{\rm E})}
={2\beta+\theta_{\rm E}\over \beta(\beta+\theta_{\rm E})}\;,
\ee
\be
h={\theta_{\rm E}/\theta -2(\theta_1/\theta)^2\over\theta-\theta_{\rm
    E}}={\theta_{\rm E}(\beta+\theta_{\rm E})-2\theta_1^2
\over(\theta_{\rm E}+\beta)^2\beta} \;,
\ee
for which the deformation function becomes
\be
f(\beta)={\theta_{\rm E}\over\beta}-{2\theta_1^2\ln(\beta+\theta_{\rm
  E}) \over \beta(\beta+\theta_{\rm E})}\;.
\elabel{ftarget}
\ee
This functional form of $f$ is, however, only a valid desciption for
large arguments; in particular, $f$ is not an even function of
$\beta$. We can now modify $f$ such that it retains the form
(\ref{eq:ftarget}) for large $\beta$, but becomes a function of
$\beta^2$ only. A simple way to achieve this is to replace $\beta$ in 
(\ref{eq:ftarget}) by $\sqrt{\beta^2+\beta_0^2}$, for a suitably
chosen $\beta_0$. 
\begin{figure}
\includegraphics{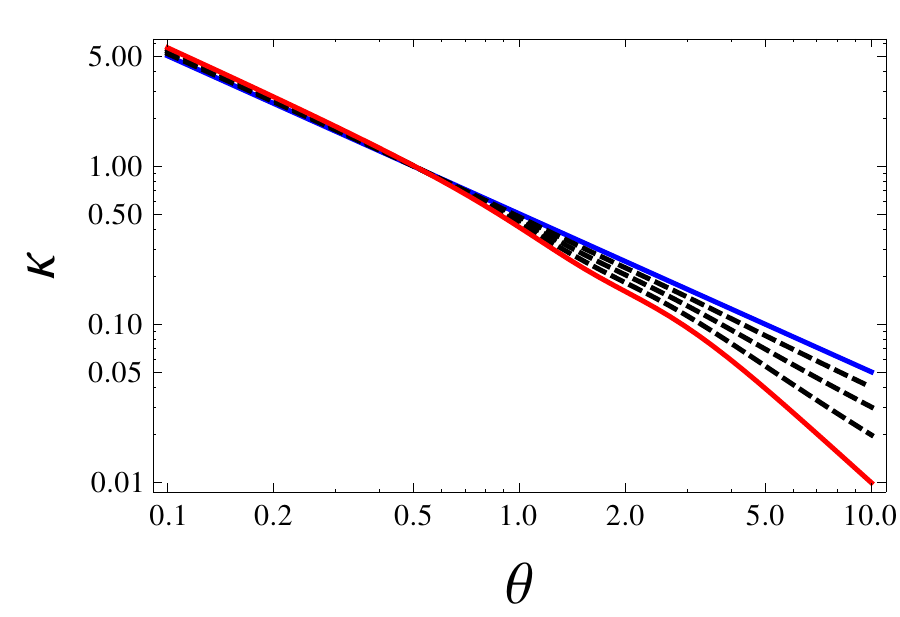}
\caption{Second example of an SPT, in the axi-symmetric case, between
  an SIS with mass distribution $\kappa(\theta)$ and other mass distributions
  $\hat\kappa(\theta)$, characterized by a deformation function
\[
f(\beta)=a\eck{{\theta_{\rm E}\over\sqrt{\beta^2+\beta_0^2}}
-{2\theta_1^2\ln\rund{\sqrt{\beta^2+\beta_0^2}
+\theta_{\rm
  E}} \over \sqrt{\beta^2+\beta_0^2}
\rund{\sqrt{\beta^2+\beta_0^2}+\theta_{\rm E}}} }\;,
\]
where we set $\theta_1=\theta_{\rm E}$, $\beta_0=3\theta_{\rm E}/2$,
and different values of $a=0,1/4,1/2,3/4,1$. For $a=0$ (blue curve),
the original SIS is obtained, for $a=1$ (red curve), $\hat\kappa$
behaves like $(\theta_1/\theta_{\rm E})^2$ for large radii}
\label{fig:Fig2}
\end{figure}
In Fig.\ts\ref{fig:Fig2}, we illustrate this example, showing that by
this choice of $f$ one finds a mass distribution $\hat\kappa$ which
has a significantly different form from $\kappa$, nevertheless giving
rise to exactly the same strong lensing properties for all source
positions. 

Note that the slope near the Einstein radius of the profiles shown in
Figs.\ts\ref{fig:Fig1} and \ref{fig:Fig2} are quite different, hence
these mass models will give rise to very different predictions for the
product of time delay and Hubble constant, $\tau=H_0\,\Delta t$, which
reinforces the point made in SS13.

\section{\llabel{Sc4}The general case}
We now drop the assumption of axi-symmetry; in this case, the matrix
$\hat\A$ in (\ref{eq:hatA}) will not be symmetric in general. This is due
to the fact that $\hat\A$ is symmetric only if the directions of the
eigenvectors of $\A$ and ${\cal B}$ are the same. For source points
which correspond to multiple images, even if we could arrange ${\cal
  B}$ to have the same eigendirections as $\A(\vc\theta_1)$, it will
not have the same eigendirections as $\A(\vc\theta_i)$ for the other
images $\vc\theta_i$. The only possibility to keep $\hat\A$ symmetric
for a general lens is to have ${\cal B}$ proportional to the unit
matrix, which is the case for $\hat{\vc\beta}=\lambda\vc\beta$, which
recovers the MST mentioned before.

The asymmetric nature of $\hat \A$, discussed in more detail in
Sect.\ts\ref{sc:Sc4.1},  
implies that the deflection law
$\hat{\vc\alpha}$, which yields exactly the same strong lensing
properties as $\vc\alpha$, cannot be derived as the gradient of a
deflection potential. Thus, in general there does not exist a surface
mass density $\hat\kappa$ which generates the same strong lensing
properties for all source positions as $\kappa$. However, it may be
possible that one can find a surface mass density which generates
almost the same mapping for all sources. In Sect.\ts\ref{sc:Sc4.2} we
obtain a crude estimate for the amplitude of the asymmetry of
$\hat\A$, for the special case of a quadrupole lens. An illustrative
example is discussed in more detail in Sect.\ts\ref{sc:Sc4.3}, where
we show explicitly that the impact of the asymmetry of $\hat\A$ can be
very small in realistic cases. 

\subsection{\llabel{Sc4.1}The transformed mass profile}
The Jacobi matrix of the original lens mapping has the form
\be
\A=\eck{1-\kappa(\vc\theta)}{\cal I} -\gamma(\vc\theta)
\left(
    \begin{array}{cc}
      \cos 2\vt & \sin 2\vt\\ 
      \sin 2\vt &-\cos 2\vt  \\
    \end{array}
  \right) \;,
\elabel{Anormal}
\ee
where ${\cal I}$ is the unit matrix. We will write the matrix ${\cal
  B}$ in a similar form,
\be
\B= B_1 {\cal I} +B_2
\left(
    \begin{array}{cc}
      \cos 2\eta & \sin 2\eta\\ 
      \sin 2\eta &-\cos 2\eta  \\
    \end{array}
  \right) \;.
\ee
The angle $\vt$ is the phase of the shear of the original lens
mapping; similarly, $\eta$ is the phase of the shear of the mapping
$\hat{\vc\beta}(\vc\beta)$ at the position $\vc\beta(\vc\theta)$. 
We then obtain
\bea
\hat\A&=&B_1(1-\kappa){\cal I}+B_2(1-\kappa)\left(
    \begin{array}{cc}
      \cos 2\eta & \sin 2\eta\\ 
      \sin 2\eta &-\cos 2\eta  \\
    \end{array}
  \right) \nonumber \\
&-&\gamma B_1 \left(
    \begin{array}{cc}
      \cos 2\vt & \sin 2\vt\\ 
      \sin 2\vt &-\cos 2\vt  \\
    \end{array}
  \right) 
\elabel{hatAbig}\\
&-&\gamma B_2 \left(
    \begin{array}{cc}
\cos 2(\eta-\vt) & -\sin 2(\eta-\vt)\\ 
      \sin 2(\eta-\vt) &\cos 2(\eta-\vt)  \\
    \end{array}
  \right) \;. \nonumber
\eea
The asymmetry of $\hat\A$ is due to the fact that these two phases are
different in general. As we will show below, in the axi-symmetric case,
these two phases are the respective polar angles of $\vc\theta$ and
$\vc\beta$; the fact that $\vc\beta$ and $\vc\theta$ are collinear in
the axi-symmetric case guarantees that these two angles are the same
(or differ by $\pi$). This then implies the symmetry of $\hat\A$.

We now define the modified surface mass density
$\hat\kappa(\vc\theta)$ through the trace of $\hat\A$, ${\rm tr}\hat\A
=2(1-\hat\kappa)$, i.e., in the same way as if the Jacobian $\hat\A$ was
derived from a deflection potential; this yields
\be
\hat\kappa=1+B_1 (\kappa-1)+B_2 \gamma \cos 2(\eta-\vt)\;,
\elabel{hatkap}
\ee
where it should be kept in mind that $\kappa$, $\gamma$ and $\vt$
depend on $\vc\theta$, and the $B_i$ and $\eta$ depend on
$\vc\beta(\vc\theta)$. In addition, we characterize the asymmetry of
$\cal\A$ by
\be
\hat\kappa_{\rm I}=B_2 \gamma \sin 2(\eta-\vt)\;.
\elabel{hkI}
\ee

We can easily show that Eq.\ts(\ref{eq:hatkap}) for $\hat\kappa$
reduces to our earlier result for an axi-symmetric mass profile. From
(\ref{eq:1D-SPT}) in the form
$\hat{\vc\beta}=\eck{1+f(|\vc\beta|)}\vc\beta$, we obtain
\be
B_1=1+f+{\beta\,f'\over 2}\; ;\quad
B_2={\beta\,f'\over 2}\;,
\elabel{B1B2}
\ee
and the phase is $\eta=\phi$, where $\phi$ is the polar angle of
$\vc\beta$. The shear of the lens is $\gamma=\kappa-\bar\kappa$, and
its phase agrees with the polar angle of $\vc\theta$, $\vt=\vp$. Here,
$\bar\kappa(\theta)=m(\theta)/\theta^2$ is the mean convergence within
radius $\theta$. Since either $\phi=\vp$ or $\phi=\vp+\pi$, the
asymmetric term in (\ref{eq:hatAbig}) vanishes, and the modified
convergence becomes
\bea
\hat\kappa&=&1+\rund{1+f+\beta\,f'/2}(\kappa-1)+(\kappa-\bar\kappa)\,\beta\,f'/2
\nonumber \\
&=&\kappa+ (\kappa-1)\,f+(2\kappa-1-\bar\kappa)\,\beta\,f'/2 \;,
\eea
which is seen to agree with (\ref{eq:kh-axi}), since $\theta \det\A
=(\theta-\alpha)(1+\bar\kappa-2\kappa)$.

One can expect that the convergence $\hat\kappa$ defined in
(\ref{eq:hatkap}) yields a deflection law which very closely
resembles that of $\hat{\vc\alpha}$ if the asymmetry of $\hat\A$, as
characterized by $\hat\kappa_I$, is small compared to $\hat\kappa$. 
This will be the case if the source plane deformation, which
determines the amplitude of $B_2$, is sufficiently small, and/or if
the misalignment between the shear $\gamma$ of the lens mapping and
that of the mapping $\hat{\vc\beta}(\vc\beta)$ is small. The latter
aspect depends on the kind of lens mapping one is dealing with.

\subsection{\llabel{Sc4.2}Example: The quadrupole lens}
As an illustrative case, we consider a quadrupole lens, i.e., an
axi-symmetric matter distribution characterized by the convergence
$\kappa(|\vc\theta|)$ plus some external shear, so that the lens
equation becomes
\be
\vc\beta=\eck{1-\bar\kappa(|\vc\theta|)}\vc\theta
-\left(
    \begin{array}{cc}
\gamma_{\rm p} & 0\\ 
      0 &   -\gamma_{\rm p}\\
    \end{array}
  \right) \vc\theta\;,
\elabel{QPLE}
\ee
where $\gamma_{\rm p}$ is a constant shear caused by some external
`perturbing' large-scale mass distribution.\footnote{In general, such
  a large-scale perturber will also induce some external
  convergence. However, for simplicity we will neglect this effect
  here, since it can be easily scaled out by a mass-sheet
  transformation \citep[see, e.g.][]{SaasFee1}.}  Denoting the shear of
the main lens by 
$\gamma_{\rm
  m}(\vc\theta)=\kappa(|\vc\theta|)-\bar\kappa(|\vc\theta|)$, the 
Jacobian reads
\be
\A=(1-\kappa){\cal I}-\left(
    \begin{array}{cc}
\gamma_{\rm p}+\gamma_{\rm m}\cos 2\vp & \gamma_{\rm m}\sin 2\vp\\ 
      \gamma_{\rm m}\sin 2\vp &  -\gamma_{\rm p}-\gamma_{\rm m}\cos 2\vp\\ 
    \end{array}
  \right) \;,
\ee
which can be compared with the form (\ref{eq:Anormal}) of $\A$; this 
yields for the shear components of $\A$ 
\bea
\cos 2\vt&=&{\gamma_{\rm p}+\gamma_{\rm m}\cos 2\vp\over\gamma}
\; ;\quad
\sin 2\vt={\gamma_{\rm m}\sin 2\vp\over\gamma} \;,
\nonumber \\
\gamma^2 &=& \gamma_{\rm p}^2+\gamma_{\rm m}^2+2\gamma_{\rm p}\gamma_{\rm
m}\,\cos2\vp  \; .
\eea
For the SPT, we again consider a radial stretching,
$\hat{\vc\beta}=\eck{1+f(|\vc\beta|)}\vc\beta$, for which the
coefficients of ${\cal B}$ are given in (\ref{eq:B1B2}), and for which
the phase $\eta$ equals the polar angle $\phi$.  We calculate
$\vc\beta$ from the lens equation (\ref{eq:QPLE}), to find
\bea
\cos 2\phi&=&{\beta_1^2-\beta_2^2\over \beta^2}
={\eck{(1-\bar\kappa)^2+\gamma_{\rm p}^2}\cos 2\vp-2\gamma_{\rm
    p}(1-\bar\kappa) \over (\beta/\theta)^2}\;, \nonumber \\
\sin 2\phi&=&{2\beta_1\beta_2\over \beta^2}
={\eck{(1-\bar\kappa)^2-\gamma_{\rm p}^2}\sin 2\vp 
\over (\beta/\theta)^2}\;,  \\
\beta^2&=&\theta^2\eck{(1-\bar\kappa)^2+\gamma_{\rm p}^2
-2(1-\bar\kappa)\gamma_{\rm p}\cos 2\vp} \;.\nonumber 
\eea
From these relations, we can calculate the products
$\gamma\,\cos 2(\eta-\vt)=\gamma\,\cos2(\phi-\vt)$
and $\gamma\,\sin2(\phi-\vt)$, which enter the
quantities (\ref{eq:hatkap}) and (\ref{eq:hkI}),
\bea
\gamma\!\!\!\!\!&&\!\!\!\!\!\cos2(\phi-\vt)
=\rund{\theta\over\beta}^2\Bigl\{ \gamma_{\rm m}(1-\bar\kappa)^2-2\gamma_{\rm p}(1-\bar\kappa)
\elabel{cosdphi}  \\
&&+\gamma_{\rm p}\eck{(1-\bar\kappa)^2-2\gamma_{\rm
    m}(1-\bar\kappa)+\gamma_{\rm p}^2}\cos 2\vp
+\gamma_{\rm m}\gamma_{\rm p}^2 \cos 4\vp \Bigr\} \;, \nonumber
\eea
\bea
\gamma\!\!\!\!\!&&\!\!\!\!\!\sin 2(\phi-\vt)
=-\rund{\theta\over\beta}^2\gamma_{\rm p}\,\Bigl\{
\gamma_{\rm m}\gamma_{\rm p}\sin(4\vp)\nonumber \\
&&+
\eck{\gamma_{\rm p}^2-(1-\bar\kappa)(2\gamma_{\rm m}+1-\bar\kappa)}
\sin2\vp  \Bigr\}
 \;.
\elabel{sindphi}
\eea
Combining (\ref{eq:hatkap}), (\ref{eq:B1B2}) and (\ref{eq:cosdphi}),
we then obtain for the transformed surface mass density
\be
\hat\kappa=(1+f)\kappa -f +{\beta\,f'\over
  2}\eck{\kappa-1+\gamma\cos2(\phi-\vt)} \;.
\ee
If we now approximate the function $f(\beta)$ near the origin by
$f(\beta)=f_0+f_2\beta^2/2$, where we accounted for the fact that
$f$ is an even function, we see that $\beta\,f'/2=f_2\beta^2/2$,
so that
\bea
\hat\kappa&=&(1+f_0)\kappa -f_0 +{f_2\,\beta^2\over2}
\eck{2\rund{\kappa-1}+\gamma\cos2(\phi-\vt)} \nonumber \\
&=&(1+f_0)\kappa -f_0 +{f_2\,\theta^2\over2}\Bigl\{
\gamma_{\rm m}\eck{2 \gamma_{\rm p}^2+3 (1-\bar\kappa)^2} \nonumber\\
&&-2(1-\bar\kappa)\eck{(1-\bar\kappa)^2+2 \gamma_{\rm p}^2}\\
&+&\eck{5\gamma_{\rm p}(1-\bar\kappa)^2-6\gamma_{\rm p}\gamma_{\rm
    m}(1-\bar\kappa)+\gamma_{\rm p}^3}\cos2\vp
+\gamma_{\rm m}\gamma_{\rm p}^2 \cos 4\vp \Bigr\} \nonumber
\eea
% 
% (kbar[t]+gm[t])(1+f0)-f0+f2 t^2/2 
% (   (5gp(1-kbar[t])^2-6 gp gm[t](1-kbar[t])+gp^3   )Cos[2p] 
% -2(1-kbar[t])((1-kbar[t])^2+2gp^2)+gm[t](2 gp^2+3(1-kbar[t])^2) 
% + gm[t] gp^2 Cos[4p]  )
The terms independent of $f_2$ present just the MST (\ref{eq:MST}),
with $\lambda=1+f_0$. The non-linear part of the SPT, here
parametrized by $f_2$, both adds a monopole contribution to
$\hat\kappa$, as well as contributions which depend on $\vp$. The
monopole contribution itself has two parts, one independent of
$\gamma_{\rm p}$ -- which can be shown to agree with (\ref{eq:kh-axi})
-- and one $\propto \gamma_{\rm p}^2$. This later contribution is
expected to be small, for reasonably small values of the external
shear. The angle-dependent contributions to $\hat\kappa$ vanish for
$\gamma_{\rm p}=0$. At the tangential critical curve of the
axi-symmetric lens, where $\bar\kappa=1$, these angle-dependent terms
are at least of order $\gamma_{\rm p}^2$, again expected to be
small. For locations away from the Einstein circle, the leading-order
terms are $\propto \gamma_{\rm p}$ and thus typically a factor of
$\gamma_{\rm p}$ smaller than the $f_2$-induced contributions to the
monopole term.

With the same form of $f(\beta)$, the quantity (\ref{eq:hkI})
describing the asymmetry of $\hat\A$ becomes
\bea
\hat\kappa_I&\approx&
-{\gamma_{\rm p}\over 2}\theta^2 f_2 \elabel{kapI}\\
&\times&
\eck{\gamma_{\rm p}^2-(1-\bar\kappa)(2\gamma_{\rm m}+1-\bar\kappa)
+2\gamma_{\rm m}\gamma_{\rm p}\cos(2\vp)}\sin{2\vp} \;.\nonumber
\eea
From this result we see that the asymmetry vanishes if
$\gamma_{\rm p}=0$. Furthermore, it vanishes on the axes,
$\vp=n\,\pi/2$, $n=0,1,2,3$, since points $\vc\theta$ on the symmetry
axes are mapped onto the corresponding axis in the source plane, so
that the shear matrices are aligned in this case.

As expected, $\hat\kappa_I$ is proportional to the curvature $f_2$
of the distortion function at the origin, and further has as prefactor
the external shear $\gamma_{\rm p}$. Both of these are typically
small; characteristic values for the external shear in strong lens
systems are $\gamma_{\rm p}\lesssim 0.1$. We have seen in
Sect.\ts\ref{sc:Sc2} that distortion functions with too large curvature
lead to unphysical mass distributions, so that $f_2\theta^2$ will
be considerably smaller than unity in the multiple image region. Hence,
the prefactor will be smaller than about $10^{-2}$. In fact, in the
example considered in the following subsection, this prefactor is of
order $2\times 10^{-4}$. Furthermore, we also point
out that the dominant term in the bracket (the middle one) will be
small near the critical curve, since the latter will be close to the
radius $\theta_{\rm E}$ of the Einstein circle of the original
axi-symmetric lens model, where $\bar\kappa=1$. Finally we note that
the $\vp$-average of $\hat\kappa_I$ vanishes, due to the sine-factor,
so that $\hat\kappa_I$ will oscillate around zero as one considers
circles of constant $\theta$. Overall, we thus expect $\hat\kappa_I$
to be small, so that the difference between the deflection angle
$\hat{\vc\alpha}(\vc\theta)$ and the one derived from the modified
surface mass density (\ref{eq:hatkap}) will be small.

\subsection{\llabel{Sc4.3}An illustrative example}
Whereas a more detailed investigation of this question will be
deferred to a later publication, we will illustrate the SPT with a
simple example. In SS13, we found that a quadrupole lens with a mass
model $\kappa(\theta)$ consisting of a Hernquist profile (representing
the baryonic mass of a lens galaxy) and a generalized NFW-profile (to
approximate the central part of the dark matter halo) is almost
degenerate with a model $\kappa_{\rm PL}(\theta)$ corresponding to a
power law with an inner core.  Since the time-delay ratios do not
simply scale by a constant factor, the transformation between these
two mass models is not a MST. Instead, as we will show next, this
transformation is an example of the SPT discussed here.

The two mass models we study hereafter are the same as in Sect. 4.2 of
SS13, namely a fiducial model composed of a spherically symmetric
Hernquist+generalised NFW with an external shear $(\gamma,
\theta_\gamma)=(0.1, 90^{\circ})$, and a target power-law model with a
core radius $\theta_c=0\arcsecf 1$, a logarithmic (three-dimensional)
slope $\gamma'= 2.24$ and an external shear $(\gamma,
\theta_\gamma)=(0.09, 90^{\circ})$. Hereafter, we note the lensing
quantities associated to the power-law model with a {\it {hat}},
e.g. $\kappa_{\rm PL} = \hat\kappa$.

\begin{figure*}
\centering
\begin{tabular}{cc}
\includegraphics[scale=0.5]{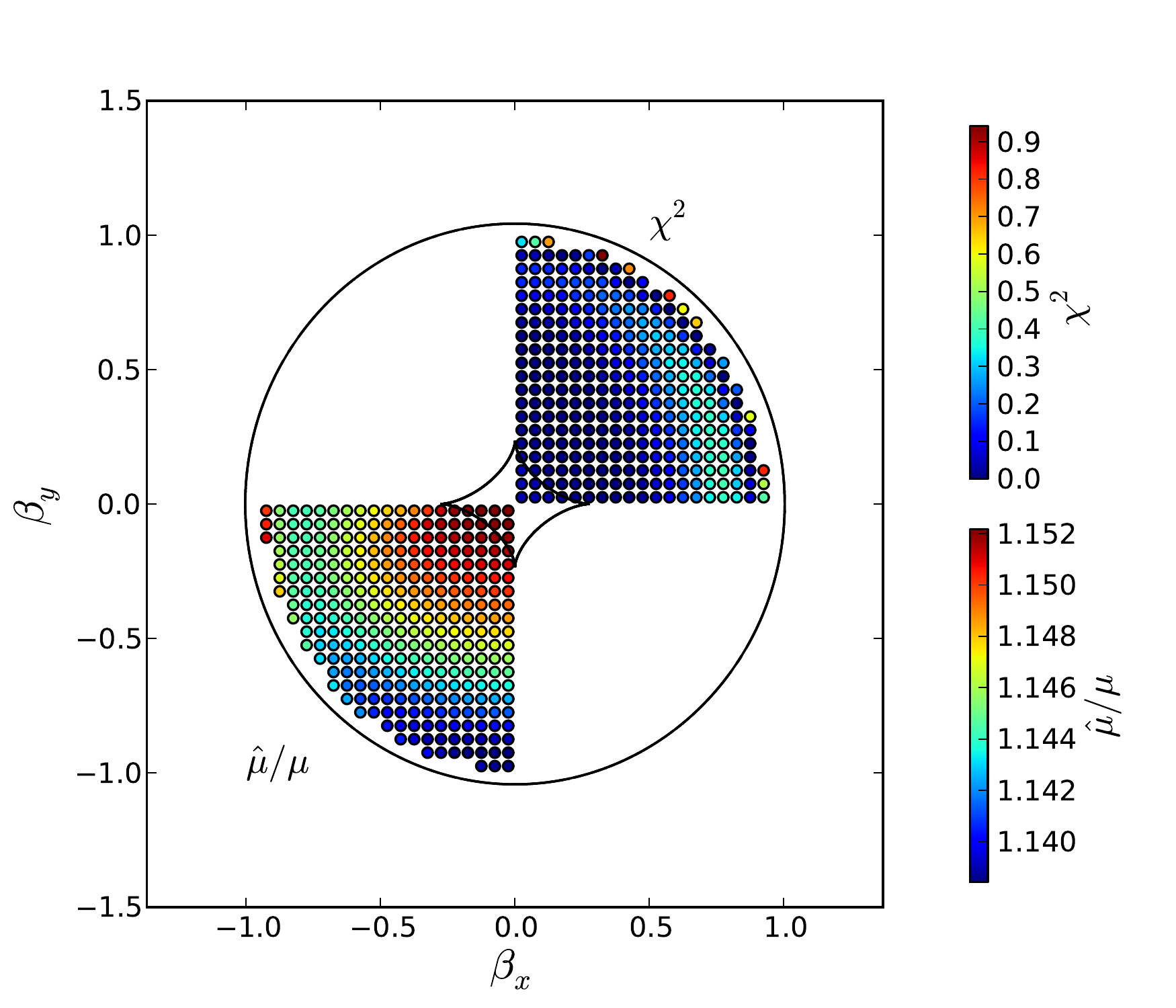} & \includegraphics[scale=0.5]{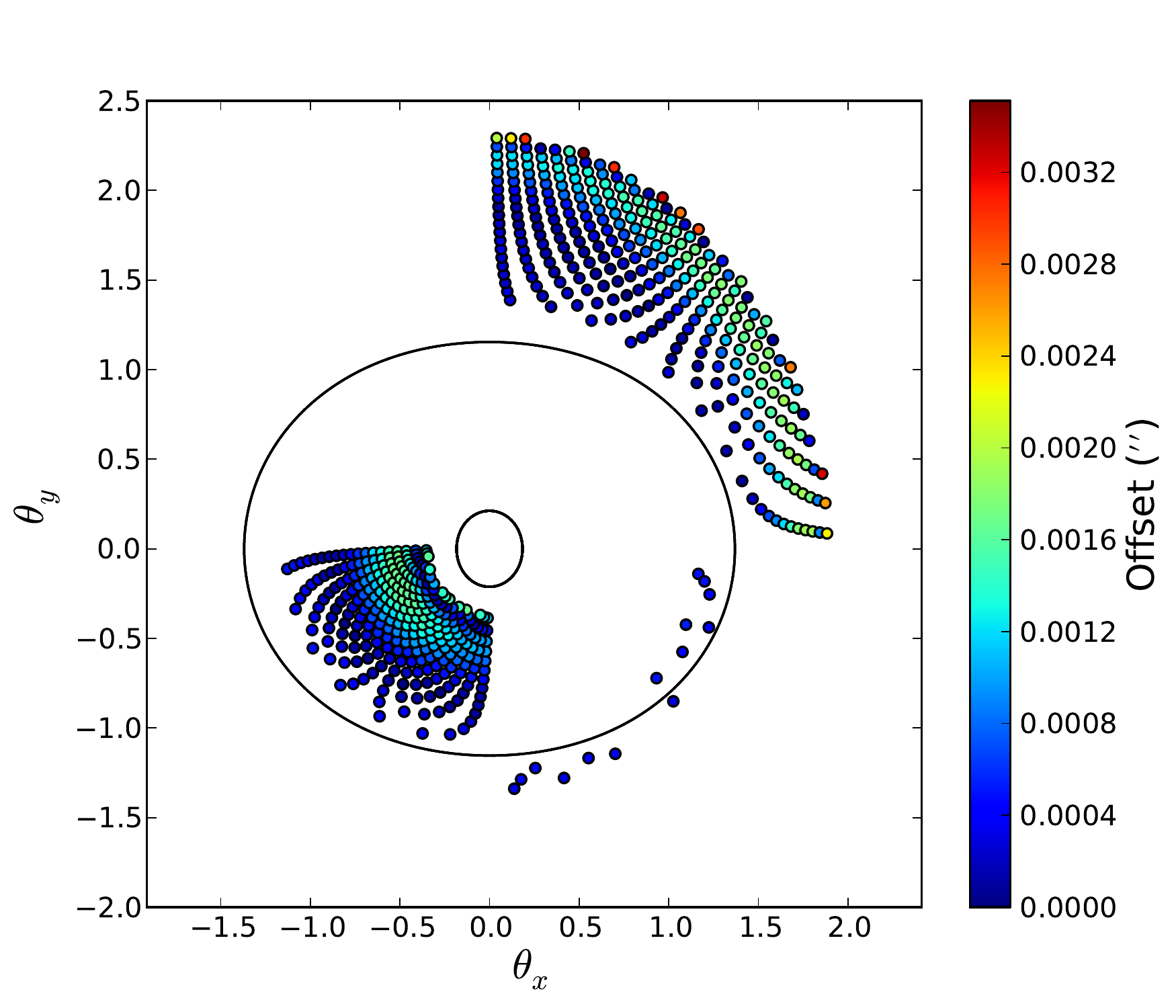} \\
\end{tabular}
\caption{{\it {Right:}} Set of mock lensed images produced by our
  fiducial lens model (using a grid of source positions shown in the
  first quadrant of the left panel). The color coding shows the offset
  in arcseconds, between the original image positions and those found
  with the power-law model. {\it {Left:}} First quadrant (upper
  right): Grid of source positions lensed with our fiducial model to
  produce the images shown in the right panel. The color coding
  indicates the $\chi^2$ associated to each source, assuming an
  astrometric uncertainty on the image position of
  $\sigma_{\theta_{x,y}} = 0.004^{\prime\prime}$. In the third
  quadrant (lower left), the sources do not have corresponding lensed
  images on the right pannel, and are colored as a function of the
  ratio of their total magnification as derived with the power-law and
  the fiducial model (i.e. $\mu_{\rm PL}/\mu_{\rm fid} = \hat
  \mu/\mu$) %Note that only the sources located in the first quadrant have corresponding image positions on the right pannel.
}
\label{fig:Fig3}
\end{figure*}

Using our fiducial mass distribution and the public lens modeling code
\texttt{lensmodel} \citep[v1.99; ][]{Keeton2001}, we create mock
images of a uniform grid of 19$\times$19 sources covering the first
quadrant of the source plane from $\vc\beta = (\beta_x, \beta_y) =
(0.025, 0.025)\arcsec$ to $(\beta_x, \beta_y) = (0.975,
0.975)\arcsec$. The sources and the mock lensed images are shown in
Fig.~\ref{fig:Fig3}. As we have shown in SS13, those lensed images are
also reproduced to a good accuracy with the power-law model, provided
that the sources are now located at positions $\hat{\vc\beta}$.  The
right panel of Fig.~\ref{fig:Fig3} shows the offset (in arcsec)
between the mock lensed images $\vc\theta$ and the lensed images
$\hat{\vc\theta}$ corresponding to the source positions
$\hat{\vc\beta}$. We see that the two sets of images are almost
identical. Assuming an astrometric accuracy of $0\arcsecf004$ in the
image plane, we have calculated an astrometric $\chi^2$ for each
source, as shown in the left panel\footnote{Note that to ease
  legibility, we have shown only those sources with $\chi^2 < 1$.} of
Fig.~\ref{fig:Fig3}. The two models are only {\it {approximately
    degenerate}} since $\chi^2$ deviates significantly from 0 in the
vicinity of the outer caustic. The figure reveals that the degeneracy
between the two models is valid for $|\beta| \lesssim 0\arcsecf9$.

\begin{figure}
\includegraphics[scale=0.45]{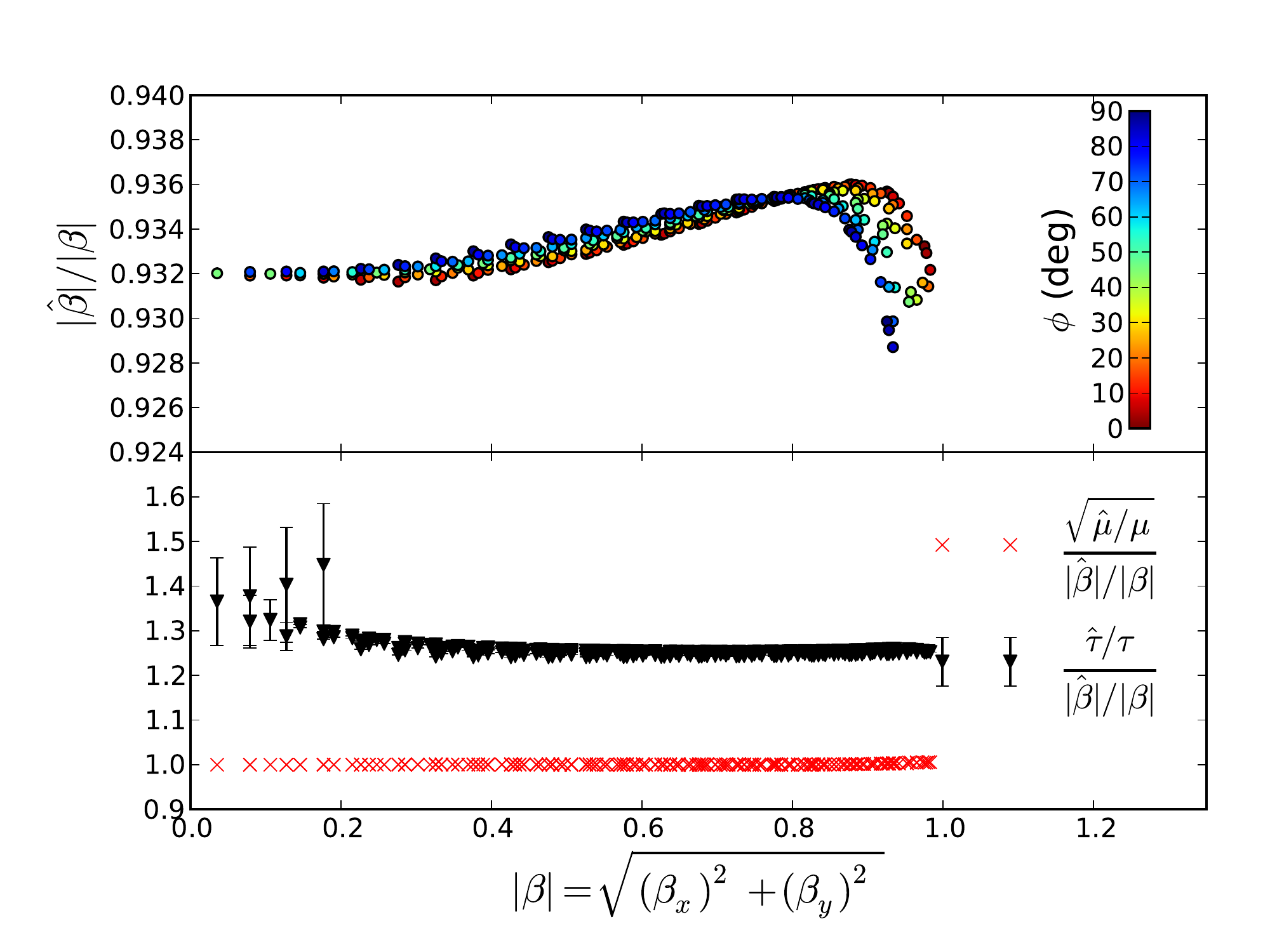}
\caption{ Properties of an approximate SPT $\vc\beta \to \hat{\vc\beta}$
  (\ref{eq:1D-SPT}) for a quadrupole lens. This SPT transforms a
  surface mass density $\kappa$ for our fiducial model into a surface
  mass density $\hat \kappa$ of a power law with a finite
  core. {\it{Top:}} Proxy to the factor $1+f(\vc\beta)$ which scales
  the positions, given by the ratio of source position modulus
  obtained for the power law and the fiducial models, i.e. $|\hat{\vc
    \beta}|/|\vc\beta|$. The color coding of this quantity as a
  function of the polar angle of the source $\phi$ (in degrees) shows
  the azimuthal change of $f(\beta)$, or in other words, the slight
  anisotropy of the SPT. {\it {Bottom:}} The curves show how the SPT
  modifies the magnification $\mu \to \hat \mu$ and the time delay
  $\tau \to \hat \tau$. The red crosses show $\sqrt{\hat \mu / \mu}/
  (|\hat{\vc\beta}|/|\vc\beta|)$ and the black triangles show $(\hat
  \tau / \tau)/(|\hat{\vc\beta}|/|\vc\beta|)$ for each source
  position. The error bars give the standard deviation of the quantity
  as obtained for the different images of the same source. The
  standard deviation is negligible for the magnification but can reach
  10\% for the time delay }
\label{fig:Fig4}
\end{figure}

To characterize the nature of the degeneracy we first consider the
relation between $\vc\beta$ and $\hat{\vc\beta}$. In case of a MST,
$\hat{\vc\beta} = \lambda\,\vc\beta$, and $\lambda$ is constant. We
show in the upper panel of Fig.~\ref{fig:Fig4} the variation of
$|\hat{\vc\beta}|/|{\vc\beta}|$ with $|\vc\beta|$ and with the polar
angle of the source, $\phi$. We see that $|\hat{\vc\beta}|/|\vc\beta|=0.932$
at the origin, and it increases monotonically to
$|\hat{\vc\beta}|/|\vc\beta|=0.936$ when $|\beta|\sim 0\arcsecf9$. 
This implies that for this specific example, the product $\theta^2
f_2$ occurring in (\ref{eq:kapI}) is about $4\times 10^{-3}$, yielding
an asymmetry of the Jacobian $\hat\A$ of less than $\sim 2\times
10^{-4}$.

The sharp decrease of the function $(1+f)$ for $|\vc\beta|\gtrsim 0\arcsecf9$
likely reflects the approximate character of the degeneracy between
$\kappa$ and $\hat\kappa$.\footnote{We point out here that there was
  no fine-tuning involved in constructing this special example, i.e.,
  this example from SS13 turned out to be almost degenerate with a
  cored power-law mass profile. The fact that it almost corresponds to
  an SPT is accidental. Some fine-tuning of the mass profiles,
  equivalent to a fine-tuning of the function
  $\hat{\vc\beta}(\vc\beta)$, would enable an even better agreement
  between the lensing properties of the two models.}  The change of
$|\hat{\vc\beta}|/|{\vc\beta}|$ with $\vc\beta$ demonstrates that the
transformation is not a MST. On the other hand, the small dependence
of $|\hat{\vc\beta}|/|\vc\beta|$ on $\phi$ shows that the mapping
$\vc\beta \to \hat{\vc\beta}$ is almost isotropic, but not exactly so.

We now compare the magnification of the lensed images and the
corresponding total magnification $|\mu|$ of the source. The spatial
variation of the magnification ratio in the source plane is displayed in
the third quadrant of Fig.~\ref{fig:Fig3}. We see that the factor by
which the magnifications are transformed is the square of the factor
transforming the positions, as shown on the bottom panel of
Fig.~\ref{fig:Fig4} which shows $\sqrt{|\mu|/|\hat\mu|}$ normalised
by $|\vc\beta|/|\hat{\vc\beta}|$.  In other words, if $\hat{\vc\beta} =
[1+f(\vc\beta)]\,\vc\beta$, the magnification has to be transformed as
$\hat\mu = [1+f(\vc\beta)]^{2}\,\mu$. This behaviour is similar to
what is observed for the MST, except that there is a dependence on
$\vc\beta$.

Finally, we display also on Fig.~\ref{fig:Fig4} the change of the time
delay $\tau/{\hat\tau}$ normalised by
$|\vc\beta|/|\hat{\vc\beta}|$. While the delays should scale like the
source positions in case of a MST, they do not so for the SPT. The
ratio ($\tau/\hat\tau)/(|\vc\beta|/|\hat{\vc\beta}|)$ is almost
constant but differs significantly from 1. In addition, we see that
for sources interior to the inner (astroid) caustic, the time delay
ratios between lensed images (of the same source) are not
conserved. The error bars in Fig.~\ref{fig:Fig4} show, for each
source, the spread of $\tau/\hat\tau$ among the lensed images. This
spread may reach 10\% for some peculiar source positions, but
generally arises from the deviation of only one of the lensed
images. Owing to accuracies on the time delays of a few percents
currently achieved for quads \citep[e.g.][]{Fassnacht2002,
  Courbin2011, Tewes2012}, time delay measurements in quadruply lensed
systems could in the most favourable cases break the SPT independently
of $H_0$.

In summary, we have shown that the degeneracy between a composite and
a power-law model with finite core identified in SS13 is an example of
an approximate SPT in the case of a quadrupole lens. The non-uniform
rescaling of the source morphology implies also a rescaling of its
surface brightness. Although this means that in principle an
examination of the source should allow one to exclude some inadequate
degenerate models, it is likely that, as for the example shown here,
the source corresponding to the two different models will both have
similarly plausible morphologies and surface brightnesses. Again,
time delay ratio measurement may play a critical role to break the
degeneracy between SPT-generated models. First, because the time delay
ratios are not perfectly conserved when 3 time-delays of the same
source are observed. Second, because the time delay is not invariant
under an SPT, and so, if the value of the Hubble constant is assumed
to be known from other observations, the degeneracy may be at least
partially broken. Finally, we also note that the amplitude of the
shear also gets modified by the SPT.

% Fig generated with fitpwl.figcaustSPT(fid = 'SPT_fiducial2_wc_g3/Optim02_fix_fid_wc_g3_fix.dat', alph='SPT_fiducial2_wc_g3/Optim02_alph_fid_wc_g3_fix.dat', crit='SPT_fiducial2_wc_g3/crit_fid_wc_g3_fix.crit', chitresh=1.0)

% Fig. generated with fitpwl.figfbeta(mfile='SPT_fiducial2_wc_g3/fbeta.txt')

\section{\llabel{Sc5}Discussion and conclusions}
We have shown that there exists a transformation of the deflection
angle in gravitational lens systems which (1) leaves all strong
lensing observables invariant, except the time delay, and (2) is much
more general than the well-known mass-sheet transformation. For the
axi-symmetric case, this new source-position transformation is exact,
in the sense that there exists a surface mass density
$\hat\kappa(\theta)$ which corresponds to the transformed deflection
law $\hat\alpha(\theta)$. In the general case, the transformed
deflection $\hat{\vc\alpha}$ has a curl component, which causes the
resulting Jacobi matrix to attain an asymmetric contribution. 

We have then defined the convergence $\hat\kappa$ of the transformed
lens in terms of the trace of the transformed Jacobi matrix
$\hat\A$. The deflection angle obtained from this convergence is
expected to closely approximate the transformed deflection angle
$\hat{\vc\alpha}$, provided the asymmetry of $\hat\A$ is sufficiently
small. Hence, in this case the mass distribution $\hat\kappa$ will
yield almost the same strong lensing predictions as the original mass
profile $\kappa$. 

Such an approximate agreement between the original and the transformed
lensing properties is sufficient for the typical strong lens systems.
These are usually modeled by `simple' mass profiles with a small
number of parameters which fit the observed image configuration
remarkably well. Exact fitting is not required, though, for at least
two reasons: first, the observed image positions (and the observed
brightness profile for extended images) have an observational
uncertainty, which for the best optical imaging available (with HST)
is of the order of $\sim 1/10$ of a pixel, i.e., $\sim 5\,{\rm mas}$,
corresponding to $\Delta\theta/\theta_{\rm E}\sim 5\times 10^{-3}$ for
galaxies as lenses. Higher accuracy in lens modeling is thus currently
not required. Second, real mass distributions are not really smooth,
but contain substructures; these substructures are almost certainly
responsible for the mismatch between observed flux ratios of images,
and the magnification ratios obtained from simple (i.e., smooth) lens
models \cite[e.g.,][]{MS98,KochDal04,Brad04,Dandan10}. In addition,
low-mass halos along the line-of-sight to the source may change
magnifications \cite[e.g.,][]{Metcalf05,Dandan12}.  For these reasons,
flux ratios are usually not used as constraints in lens modeling. The
upper mass end of the substructure can in addition cause positional
shifts of individual images; such astrometric distortions have been
observed in several lens systems where the substructure was indeed
identified (e.g., MG\ts 0414+0534 -- e.g., Trotter et al.\
\citeyear{TWH00}; Ros et al.\ \citeyear{Ros00} -- and MG\ts 2016+112
-- e.g., Koopmans et al.\ \citeyear{KGB02}). Thus, independent of the
accuracy of the observed image positions, for physical reasons one may
not expect to reproduce the observed position to better than a few
milliarcseconds with a smooth mass model. Hence, as long as the
transformed deflection angle $\hat{\vc\alpha}$ (which yields exactly
the same strong lensing properties as the original lens) and the
deflection obtained from $\hat\kappa$ differ by less than the smallest
angular scale on which modeling by a smooth mass distribution is still
meaningful, this difference is of no practical relevance.

In the near future, images of structured Einstein rings may be
observed at higher resolution (i.e. $0\arcsecf001$) with ALMA
\citep{Hezaveh2013a}, the Extremely Large Telescope (ELT) or the James
Webb Space Telescope (JWST). It still has to be investigated how
critical could the SPT be for sources observed at those resolutions if
substructures are not explicitly included in the lens models and/or if
reasonable freedom is allowed regarding the angular structure of the
lens \citep[e.g.][]{Evans2003, Saha2006c}. Nevertheless, the expected
substructure and line-of-sight inhomogeneities mentioned above will
put a lower limit to the positional accuracy at which lens systems can
be modeled with smooth matter distributions. In any case, since the
SPT is more general than the MST, it will limit the accuracy of some
applications of strong gravitational lensing such as the detection of
substructures via the time-delay method \citep{Keeton2009}, the
estimate of $H_0$ for systems where only one time delay is measured
\citep[e.g.][]{Vuissoz2007, Suyu2012}, or the determination of the profile of
the dark matter distribution based only on strong lensing \citep[e.g.]
[]{Cohn2001, Suyu2012b, Eichner2012}. Free-form
  lens modeling might already give a hint of the impact of the SPT on
  some of these applications \citep[][]{Coles2008, Paraficz2010,
    Leier2011}.

It should also be pointed out that `simple' mass models of strong
lensing clusters typically fail to reproduce the location of multiple
images at the level of $\sim 1''$, even if the contributions of the
cluster galaxies are explicitly accounted for. This level of mismatch
is typically not considered to be a problem, since one expects the total
mass distribution of clusters to be more complicated than that of one
or a few large-scale mass components plus the mass profiles of cluster
galaxies (for which simple scalings between luminosity and mass
properties are employed). Here, the relative mismatch is
$\Delta\theta/\theta_{\rm E}\sim 1/20$, i.e., larger than for
galaxies, and correspondingly, the required agreement between
$\hat{\vc\alpha}$ and the deflection from $\hat\kappa$ is less
stringent. 

Free-form lens models developed for studying strongly lensed sources
by galaxies or clusters \citep[e.g.][]{Saha2004, Diego2005a,
  Liesenborgs2006, Coe08} fit perfectly the lensing observables and
derive ensemble of models reproducing existing data, for sets of
individual multiply-imaged sources.  Those techniques effectively
explore degeneracies between lens models \citep{Saha2001, Saha2006c},
but with the drawback that many non-physical models (e.g., dynamically
unstable or with arbitrary substructure) can be obtained. As
emphasized by Coe et al.\ (\citeyear{Coe08}), there is no unambiguous
set of criteria to define a priori whether a model is physical or
not. Hence, free-form modeling may sometimes worsen the impact
of degeneracies by exploring a parameter space which is too
large. More work is surely needed to develop schemes allowing to
explore lens degeneracies such as the SPT in a controled way, maybe
implying to combine model-based and model-free approaches, or include
perturbations around local solutions \citep{Alard2009}.

We have illustrated the behavior of the SPT for a quadrupole lens. As
an example, we have shown that in the presence of an external shear
$\gamma \sim 0.1$, a mass model constituted of a baryonic component
(modeled as a Hernquist profile) and of a dark matter component
(modeled as a generalized NFW) could be transformed into a power-law
model with a finite core. In fact, as mentioned before, this example
was not constructed as an SPT, but the (almost perfect) degeneracy
between these two mass models was found `experimentally' in SS13. We
have shown here that this degeneracy nevertheless can be traced back
to the SPT; in particular, we saw that the SPT corresponding to that
transformation is a spatially varying (at less than a percent level)
and nearly isotropic contraction of the source plane positions,
i.e. $\hat{\vc\beta} = [1+f(\vc\beta)]\,\vc\beta$. We have shown that
the magnifications are transformed such that $\hat\mu =
[1+f(\vc\beta)]^2\,\mu$, while the time delay transforms
differently. In addition, the time delay ratios are not conserved when
four lensed images are formed.

Throughout this paper we have considered strong lensing only. The weak
lensing properties are not invariant under the SPT, since it changes
the weak lensing observable, i.e., the reduced shear
$\gamma/(1-\kappa)$, except for the special case of a MST
\cite{SchnSeitz95}. Thus, the SPT is not an invariance transformation
for weak lensing studies. In the strong lensing regime, if
magnification information can be obtained from observations, the
invariance with respect to the SPT can also be broken. The latter most
likely is more relevant for clusters, where magnification can be
estimated from the observed number density of background sources,
though these estimates have a considerable uncertainty within the
strong lensing regime of clusters.

In a future work, we aim to quantify the consequences of the asymmetry
of $\hat\A$ in more detail, and thus to find criteria to derive which kind of
SPTs are allowed for a given tolerance in the changes of image
positions.

\begin{acknowledgements}
  Part of this work was
  supported by the German \emph{Deut\-sche
    For\-schungs\-ge\-mein\-schaft, DFG\/} project number SL172/1-1.
\end{acknowledgements}

\bibliographystyle{aa}
%\bibliography{/Users/sluse/work/articles/bibds}
\bibliography{MSDbib}

%\begin{thebibliography}{}

%\end{thebibliography}

\end{document}